\documentclass[11pt]{iopart}%
\usepackage{iopams}
\usepackage{graphicx}
\usepackage{graphics}
\usepackage{dcolumn}
\usepackage{bm}
\usepackage{color}
\usepackage{amsfonts}
\usepackage{amssymb}%
\usepackage{multirow}%
\providecommand{\U}[1]{\protect\rule{.1in}{.1in}}
\begin{document}

\title[Thermodynamics of metabolic energy conversion]{Thermodynamics of metabolic energy conversion under muscle load}

\author{Christophe Goupil}
\address{Laboratoire Interdisciplinaire des Energies de Demain (LIED), CNRS UMR 8236, Universit\'e Paris Diderot, 5 Rue Thomas Mann, 75013 Paris, France}
\author{Henni Ouerdane}
\address{Center for Energy Science and Technology, Skolkovo Institute of Science and Technology, 3 Nobel Street, Skolkovo, Moscow Region 121205, Russia}
\author{Eric Herbert}
\address{Laboratoire Interdisciplinaire des Energies de Demain (LIED), CNRS UMR 8236, Universit\'e Paris Diderot, 5 Rue Thomas Mann, 75013 Paris, France}
\author{Yves D'Angelo}
\address{Laboratoire Math\'ematiques \& Interactions
  J. A. Dieudonn\'e, Universit\'e C\^ote d'Azur, UNS, CNRS UMR 7351, Parc Valrose, 06108 Nice, France}
\author{Cl\'emence Goupil}
\address{H\^otel-Dieu Hospital, Assistance Publique/H\^opitaux de
  Paris, 1 place du Parvis Notre--Dame, 75004 Paris, France}
\ead{corresponding author christophe.goupil@univ-paris-diderot.fr}

%\affil[+]{these authors contributed equally to this work}
\date{\today}

\begin{abstract}
The metabolic processes complexity is at the heart of energy conversion in
living organisms and forms a huge obstacle to develop tractable thermodynamic
metabolism models. By raising our analysis to a higher level of abstraction,
we develop a compact --- i.e. relying on a reduced set of parameters ---
thermodynamic model of metabolism, in order to analyze the
chemical-to-mechanical energy conversion under muscle load, and give a
thermodynamic ground to Hill's seminal muscular operational response model.
Living organisms are viewed as dynamical systems experiencing a feedback loop 
in the sense that they can be considered as thermodynamic systems subjected to
mixed boundary conditions, coupling both potentials and fluxes.
Starting from a rigorous derivation of generalized thermoelastic and transport 
coefficients, leading to the definition of a metabolic figure of merit, we establish
the expression of the chemical-mechanical coupling, and specify the nature of the
dissipative mechanism and the so called figure of merit. The particular nature
of the boundary conditions of such a system reveals the presence of a 
feedback resistance, representing an active parameter, which is crucial for the proper 
interpretation of the muscle response under effort in the framework of Hill's model. We also
develop an exergy analysis of the so-called maximum power principle, here
understood as a particular configuration of an out-of-equilibrium system, with
no supplemental extremal principle involved.

\end{abstract}
\maketitle

\section{Introduction}

Thermodynamics provides the proper framework to describe and analyse the rich
variety of existing sources of energy and the processes allowing its
conversion from one form to another. Yet, of all known energy converters,
man--made or not, living organisms still represent a formidable challenge in
terms of thermodynamic modeling, due to their high complexity which far
exceeds that of any other artificial or natural system
\cite{Prigogine69,Prigogine71,Haynie2001,Bromberg2011,Demirel2014,Lucia2015,Bejan2017a,Bejan2017}%
. Energy conversion in living bodies is driven by metabolism, which ensures
through chemical reactions at the cellular level, and along with other vital
functions, the provision of heat necessary to maintain a normal body
temperature, as well as, to a lesser extent, the energy required for muscular
effort and motion. In the present work, we are particularly interested in
catabolism, i.e. the energy-releasing process of breaking down complex
molecules into simpler ones, and especially how, from a nonequilibrium
thermodynamics viewpoint, the energy provided by the breaking of the digested
food substances is made available to muscles for production of mechanical
effort through contraction, as described in Hill's model \cite{Hill1938,Hill1964}. 
The vision of Hill is summarized in the first statement of his Nobel Lecture in 
1922:\\

 ``\emph{In investigating the mechanism involved in the activity of striated muscle 
two points must be borne in mind, firstly, that the mechanism, whatever it be, exists 
separately inside each individual fibre, and secondly, that this fibre is in principle 
an isothermal machine, i.e. working practically at a constant
temperature.}'' \cite{HillNobel} \\

Living organisms are open nonequilibrium and dissipative systems as they
continuously exchange energy and matter with their environment
\cite{Prigogine69,Prigogine71}. Unlike classical thermodynamic engines, for
which equilibrium models may be built using extremal principles, no such a
possibility exists for the case of living organisms because of the absence of
identifiable genuine equilibrium states. Nevertheless, assuming a global
system close to equilibrium, the development of a tractable thermodynamic
model of metabolism may rely on notions pertaining to classical equilibrium
thermodynamics: on the one hand, the working fluid acting as the conversion
medium, and on the other hand, the characterization of its thermoelastic
properties. Despite the complex features of biological systems, the
identification of the nonequilibrium processes driving the transformation of
the digested food chemical potential into a macroscopic form of energy made
available for muscle work may be obtained from an effective locally linearized
approach, ideally borrowing from the phenomenological approach to
non-equilibrium thermodynamics developed by Onsager \cite{Onsager1931}.
Application of Onsager's approach and its integration on macroscopic systems make it
possible to describe the behaviour of some thermodynamic conversion machines
placed under mixed boundary conditions \cite{Apertet2013,Apertet2012a,Ouerdane2015b}, 
which cause feedback effects and the emergence of complex dynamic behaviours \cite{Goupil2016}. 
We propose here to apply this approach 
%to the case 
%of chemical conversion machines that are 
%iving organisms. 
to the case of living organisms viewed foremost as chemical conversion machines. 
In this article we thus derive first the macroscopic response from the 
local Onsager response. We then use the obtained results to compare the predictions of the model 
with the emblematic case of the muscle response proposed by Hill.

Before we proceed with the thermodynamics of metabolism, it is useful to first
clarify what we mean by energy conversion in a standard thermodynamic engine.
When energy is transferred to a system, its response manifests itself on the
microscopic level through the excitation of its individual degrees of freedom
and also on the global level whenever collective excitations are possible.
In generic terms, a thermodynamic machine is the site of the conversion of a 
``dispersed'' incident energy flow into an ``aggregated'' energy flow and a loss flow. 
This conversion is ensured by a thermodynamic working fluid whose state equations lead 
to coupling between the respective potentials. In the case of thermal machines, the dispersed 
form of energy is called heat, and the potential associated with it is temperature, while the 
aggregated form is called work and the potential associated with it is, for example, the pressure. 
Temperature and pressure are linked by one or more equations of state. The response of the fluid 
proceeds from the collective response of the working fluid's microscopic degrees of freedom.
Hence, part of the energy received by the working fluid may be made available on a global scale 
to a load for a given purpose as useful work, the rest of it being redistributed (dispersed) on 
the microscopic level, and dissipated because of internal friction and any other
dispersion process imposed by boundary conditions \cite{Ouerdane2015b}. The
conversion efficiency is thus tightly related to the share of the energy
allocated to the collective modes of the system. Metabolism differs from
energy conversion in standard heat engines in the sense that dissipation
cannot be seen as a mere waste since, in biological processes, the dispersion
of energy is, rather than solely the production of heat, a production of
secondary metabolites \cite{Demain2000}. Nevertheless, for the purpose of
describing a short duration metabolic effort, the production of dispersed
energy can be considered as a ``waste''.

The main objective of the present work is the development of a thermodynamic
description of the out-of-equilibrium steady-state biological
chemical-to-mechanical energy conversion process, in the conditions of the
production of a muscular mechanical effort of moderate duration. There are already 
a number of reviews concerning the study of muscle from an energy viewpoint 
\cite{Curtin1978,Homsher1978,Kushmerick1983}. The particularity of our approach is 
to build the model from the first and second principles, from the perspective of a conversion 
machine. To this aim, we focus on uncovering the essential features of this process considering a
basic power converter model system of incoming dispersed power (the chemical
energy flux) to aggregated macroscopic (the mechanical power) energy flux.
This conversion zone is connected to two reservoirs of chemical energy,
respectively denoted source and sink. It is the zone where energy and matter fluxes are
coupled and actual ``dispersed-to-aggregated'' conversion 
occurs. The connection between the converter and the reservoirs produces energy dispersion due to the 
resistive coupling. 

 Usually, to analyse the metabolic machine operating process in a realistic configuration --- hence
addressing the whole biological system, one needs to successively: i) express
the local energy budget; ii) integrate the local expressions over the spatial
variables; iii) explicitly include the boundary conditions. While
thermodynamic models of biological systems have been extensively developed
\cite{Prigogine69,Prigogine71,Haynie2001,Bromberg2011,Demirel2014,Lucia2015,Bejan2017a,Bejan2017,Blumenfeld1981,Volkenshtein1983,Jusup2017}, 
including the thermodynamic network approach and bond-graph methods
\cite{Atlan1973,Oster1973}, a proper account of the above three points is
often missing, leading to an incomplete thermodynamic description of the
biological energy conversion. Our abstract approach, on the contrary,
facilitates the treatment of these quite concrete aspects.

The main results of the present work are hence three-fold:

\begin{itemize}
\item the modeling and analysis of
metabolism with, in particular, the introduction and computation of effective
thermoelastic and transport coefficients.
Feeding these coefficients to proper expression of the chemical/mechanical  coupling gives rise to a dissipative mechanism. 
% using these coefficients for expressing the
%chemical/mechanical coupling gives rise to a dissipative mechanism and the
%resulting central
The resulting central physical quantity to be considered is then the
\textit{feedback resistance}, an active parameter;

\item the subsequent first-principles thermodynamic re-interpretation of
Hill's muscle model, in particular the ability to capture the velocity dependence of
the so-called \textit{extra heat coefficient} $a$;

\item the exergy-based interpretation within the present proposed approach of
the so-called ``maximum power principle'', a moot extremal principle
frequently invoked in biology, that is here understood as a particular
configuration of an out-of-equilibrium system, and hence should \textit{not}
be considered as a fundamental --- extremal --- principle.
\end{itemize}

The article is organized as follows. In Section \ref{sec2}, we recall some of
the basic thermodynamic concepts related to the working fluid properties and
the close-to-equilibrium force-flux formalism. We discuss the assumption of
linearity, defined on the local level, on which we base the model development.
We then specify our approach to biological energy conversion. The full
development of the thermodynamic model of metabolism is presented in Section
\ref{sec3}. Some of the insights that our thermodynamic model of metabolism
can offer are discussed in Sec. \ref{sec4}, where we provide a full
thermodynamic interpretation of Hill's theory of muscle loading
\cite{Hill1938}. Section \ref{sec5} goes even further and presents a brief
exergy analysis of the so-called concept of maximal power principle (MPP) in
biology, in the frame of our approach. Concluding remarks are made in Section \ref{sec6}.

\section{Basic concepts}

\label{sec2}

\subsection{Thermoelastic coefficients}

In terms of thermodynamic variables for a standard thermodynamic engine,
entropy $S$ and temperature $T$ are associated with the energy dispersion
processes because of their direct link to heat, while pressure $P$ and volume
$V$ (or other variables like voltage, electric charge, chemical potential,
particle numbers) may be associated with the energy aggregation processes,
because of their direct link to work. In a more general framework, we may
define a set of coupled variables for dispersive processes $(\Pi_{m},m)$, and
aggregative processes $(\Pi_{M},M)$. The local formulation of the Gibbs
relation that reads in standard notations $\mathrm{d}U=T\mathrm{d}%
S-P\mathrm{d}V$, with $U$ being the internal energy, can be generalized as
follows:
\begin{equation}
\mathrm{d}U=\Pi_{m}\mathrm{d}m+\Pi_{M}\mathrm{d}M \label{dU}%
\end{equation}
\noindent where $m$ and $M$ are extensive variables, respectively standing for
the dispersive and aggregative processes, and the thermodynamic potentials
$\Pi_{m}$ and $\Pi_{M}$ are their intensive conjugate variables. The $\Pi
_{m}\mathrm{d}m$ term is the generic mathematical expression for the
dispersion of the energy,
while the $\Pi_{M}\mathrm{d}M$ term represents the aggregative
process. Although Eq.~(\ref{dU}) merely
represents a higher level of description of the thermodynamic system in the
sense that the variables $m$ and $M$ are respectively assigned a dispersive
and aggregative character, it provides a quite convenient starting point for
the study of an equivalent working fluid, whose
properties are defined by equations of state. For instance, in a heat engine
that involves the coupled transport of energy and matter, $M$ would
classically correspond to matter as it characterizes a collective modality in
the energy conversion process, and $m$ would characterize the dispersed energy
and the related entropy variation. The only physical constraint in the
definition of the variables $m$ and $M$ is that their products with their
intensive conjugate variables, $m\Pi_{m}$ and $M\Pi_{M}$, have the dimension
of an energy. This aspect is completely in line with Carath\'{e}odory's
axiomatic formulation of thermodynamics based on the properties of Pfaff's
differential forms \cite{Pogliani}. In a metabolic description, the
thermoelastic coefficients define the conversion ratios and energy capacities
for the dispersed energy as reported in Table~\ref{table1}. \begin{table}[th]
\centering
{\scriptsize
\begin{tabular}
[c]{lll}\hline\hline
~ & ~ & ~\\
Thermoelastic coefficients~ & ~ Effective metabolic equivalent ~ & ~ Standard
working fluid\\\hline
~ & ~ & ~\\
${\displaystyle \beta=\frac{1}{M}\left(  \frac{\partial M}{\partial\Pi_{m}%
}\right)  _{\Pi_{M}}}$ ~ & ~ isochemical stiffness coefficient ~ & ~ constant
pressure thermal dilatation\\
~ & ~ & ~\\
${\displaystyle \chi_{\Pi_{m}}=\frac{1}{M}\left(  \frac{\partial M}%
{\partial\Pi_{M}}\right)  _{\Pi_{m}}}$ ~ & ~ constant force stretch
coefficient ~ & ~ isothermal compressibility\\
~ & ~ & ~\\
${\displaystyle C_{\Pi_{M}}=\frac{\Pi_{m}}{M}\left(  \frac{\partial
m}{\partial\Pi_{m}}\right)  _{\Pi_{M}}}$ ~ & ~ constant force dispersion
capacity ~ & ~ specific heat at constant pressure\\
~ & ~ & ~\\
${\displaystyle C_{M}=\frac{\Pi_{m}}{M}\left(  \frac{\partial m}{\partial
\Pi_{m}}\right)  _{M}}$ ~ & ~ constant deformation dispersion capacity ~ & ~
specific heat at constant volume\\
~ & ~ & ~\\\hline\hline
\end{tabular}
}\caption{Generalized thermoelastic coefficients: expressions and metabolic
definitions. For the sake of clarity, the analogues for standard working
fluids are given.}%
\label{table1}%
\end{table}The thermodynamic characterization of the coupling between the
energy conversion zone and the muscles, which are the engine equivalent, is of
paramount importance to understand how muscular effort is achieved under
different situations. Let us now introduce the coupling coefficient $\alpha$:
\begin{equation}
\alpha=-\left.  \frac{\partial\Pi_{M}}{\partial\Pi_{m}}\right)_{M}\label{alpha} 
\end{equation}
\noindent defined as the ratio of the thermodynamic potentials $\Pi_{m}$ and
$\Pi_{M}$ derivatives. This provides a quantitative means to evaluate the energy
conversion efficiency related to the aggregative process.
Note that $\alpha$ is reminiscent of the so-called entropy per particle
introduced by Callen in the context of thermoelectricity \cite{Callen1}, and
by extension, it may be seen as a direct measure of the entropy per unit of
working fluid. This coupling coefficient is related to the thermoelastic
coefficients (see Table \ref{table1}) and can be explicitly derived, provided
the working fluid equation of state $f\left(  U,\Pi_{M},\Pi_{m},M,m\right)
=0$, which relates the internal energy $U$ of the system to its thermodynamic
variables, is known. Indeed, the heat capacity ratio $\gamma=C_{\Pi_{M}}/C_{M}$ 
yields the analogue of the classical isentropic expansion factor, a
measure of the ``quality'' of energy conversion in the sense that the system tends 
to minimize dissipation. It is given by:
\begin{equation}
\gamma=\frac{C_{\Pi_{M}}}{C_{M}}=\left(  1+\frac{\beta^{2}}{\chi_{\Pi_{m}%
}^{\phantom {-1}}C_{M}}\Pi_{m}\right)  =\left(  1+\frac{\alpha^{2}\chi
_{\Pi_{m}}^{\phantom {-1}}}{C_{M}}\Pi_{m}\right)  \label{kappa}%
\end{equation}
\noindent In Eq.~(\ref{kappa}), we made use of the extended Maxwell relations
to link the thermoelastic coefficient $\beta$ to the coupling coefficient
$\alpha$, as done in Refs. \cite{Ouerdane2015,Benenti2016}:
\begin{equation}
\beta=-\frac{1}{M}\left(  \frac{\partial M}{\partial\Pi_{m}}\right)  _{\Pi
_{m}}\left(  \frac{\partial\Pi_{M}}{\partial\Pi_{m}}\right)  _{M}=-\chi
_{\Pi_{m}}^{\phantom {-1}}\left(  \frac{\partial\Pi_{M}}{\partial\Pi_{m}%
}\right)  _{M} \label{betachi0}%
\end{equation}
\noindent so that $\beta=\alpha\chi_{\Pi_{m}}^{\phantom {-1}}$. 
From the definitions above, we obtain the dimensionless quantity:
\begin{equation}
Z\Pi_{m}=\frac{\beta^{2}}{\chi_{\Pi_{m}}C_{M}}\Pi_{m} \label{ZPi}%
\end{equation}
\noindent the thermodynamic figure of merit, which characterizes the intrinsic
performance of conversion as it provides a direct measure of its dispersed-to-aggregated energy
conversion efficiency, just like the ratio of specific heats does for a usual
gas in a heat engine.

\subsection{Metabolic forces and fluxes}

\begin{figure}[ptbh]
\centering
\includegraphics[width=0.6\textwidth]{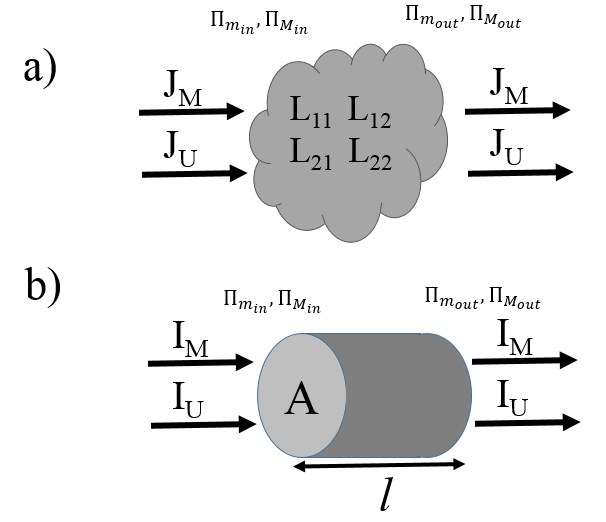}\caption{Schematic
view of a metabolic device. a) local description of the Onsager unit cell, and
b) schematic view of a portion of a living system (where $I_{U/M}=J_{U/M}A$).} 
\label{Onsager-Cell} 
\end{figure}

To develop an out-of-equilibrium description of the metabolic process, we
transpose the phenomenological linear force-flux formalism approach and
Onsager's reciprocal relations. Consider a thermodynamic unit cell (in the
general sense), able to exchange both energy and matter with its environment,
and where we can assume local equilibrium. As a matter of fact,
both the energy and matter fluxes $\mathbf{J}_{U}$ and
$\mathbf{J}_{M}$, depicted in Fig. \ref{Onsager-Cell},  are
conserved, but their coupling induces a modification of the potentials
$\Pi_{m}$ and $\Pi_{M}$ across the cell.

To account for the fluxes and the forces which derive from the thermodynamic
potentials, we extend the Gibbs relation (\ref{dU}) assuming quasi-static
working conditions:
\begin{equation}
\mathbf{J}_{U}=\Pi_{m}\mathbf{J}_{m}+\Pi_{M}\mathbf{J}_{M} \label{JU1}%
\end{equation}
\noindent where $\mathbf{J}_{U}$ represents the total energy flux; $\Pi
_{m}\mathbf{J}_{m}\equiv\mathbf{J}_{Em}$ is the dispersed microscopic energy
flux, and $\Pi_{M}\mathbf{J}_{M}$ is the aggregated energy flux, proportional
to the power produced on the macroscopic level, such as, e.g, the mechanical
power. For a living system, $\Pi_{m}$ is the chemical potential of the
digested food, which we denote $\mu$ according to standard notations, so that
the equivalent entropy flux is $\mathbf{J}_{S}=\mathbf{J}_{m}/\mu$ and
$\Pi_{M}$ is the macroscopic potential from which the muscular force derives.
The flux $\mathbf{J}_{M}$ may be seen as the metabolic intensity necessary to
maintain a given metabolic state for the living system. For our purpose, we
focus on the coupled fluxes $\mathbf{J}_{Em}$ and $\mathbf{J}_{M}$, that can
be computed using the force-flux formalism:
\begin{equation}
\left(
\begin{array}
[c]{c}%
\mathbf{J}_{M}\\
\mathbf{J}_{Em}%
\end{array}
\right)  =\left(
\begin{array}
[c]{cc}%
L_{11} & L_{12}\\
L_{21} & L_{22}%
\end{array}
\right)  \left(
\begin{array}
[c]{c}%
-\frac{1}{\Pi_{m}}\mathbf{\nabla}(\Pi_{M})\\
\mathbf{\nabla}(\frac{1}{\Pi_{m}})
\end{array}
\right)  \label{LMJem}%
\end{equation}
\noindent The off-diagonal kinetic matrix coefficients satisfy Onsager's
reciprocity relations $L_{12}=L_{21}$ \cite{Onsager1931}. In the
present case of two coupled flows, the Onsager matrix contains four terms that
are reduced to three due to the reciprocity resulting from the
Le Chatelier-Braun principle. These three parameters are respectively related
to the conductivities associated with each of the flows, on the one hand, and
the coupling coefficient between flows, on the other hand \cite{Kedem-Caplan1965}. 
For simplicity and with no loss of generality, we consider a one-dimensional description 
(in $x$-coordinate) and rewrite Eq.~(\ref{LMJem}) as:
\begin{equation}
\left(
\begin{array}
[c]{c}%
J_{M}\\
J_{Em}%
\end{array}
\right)  =\left(
\begin{array}
[c]{cc}%
L_{11} & L_{12}\\
L_{21} & L_{22}%
\end{array}
\right)  \left(
\begin{array}
[c]{c}%
\phantom{-}\frac{\displaystyle1}{\displaystyle\mu}F_{M}\\
-\frac{\displaystyle1}{\displaystyle\mu^{2}}\frac{\displaystyle\mathrm{d}\mu
}{\displaystyle\mathrm{d}x}%
\end{array}
\right)  \label{JMJem2}%
\end{equation}
\noindent where we introduced the macroscopic driving force $F_{M}%
=-\mathrm{d}\Pi_{M}/\mathrm{d}x$.

\subsection{Transport coefficients}

Let us first derive the expression of the \emph{isochemical potential
conductivity}. In this configuration, the chemical potential is constant and
the chemical energy distribution gradient within the body vanishes: $\mu$ is
constant and $J_{M}=\frac{L_{11}}{\mu}F_{M}$. The metabolic intensity $J_{M}$
is driven by the force $F_{M}=-\mathrm{d}\Pi_{M}/\mathrm{d}x$ which is nothing
but the muscular force on the macroscopic level. The metabolic conductivity is
then given by
\begin{equation}
\label{sigma}\sigma=\frac{L_{11}}{\mu}%
\end{equation}
The quantity $\sigma$ is the isochemical potential conductivity. It embodies
dissipation when the muscles produce a mechanical activity, including
locomotion but also any motionless efforts. When considering a segment of
muscle of length $l$ and section $A$, one obtains the expression of the
dissipative metabolic resistance $R_{M}=l/A\sigma$. Considering the situation
when the animal is at rest, with no mechanical activity, we now derive the
expression of the \emph{basal metabolic conductivity}. As the metabolic
intensity $J_{M}$ vanishes, the metabolic power flux $J_{Em}^{\ast}$ becomes:
\begin{equation}
J_{Em}^{*}= - \kappa_{J_{M}}\frac{\mathrm{d}\mu}{\mathrm{d}x}%
\end{equation}
\noindent where $\kappa_{J_{M}}=\frac{1}{\mu^{2}}\left[  \frac{L_{11}%
L_{22}-L_{21}L_{12}}{L_{11}}\right]  $ denotes the basal metabolic conductivity.
Since $J_{Em}^{*}$ is a measure of the consumed power density when the animal
is at rest, $\kappa_{J_{M}}$ can be considered as a metabolic conductivity
under zero load, from which we define the basal metabolic impedance: $R_{E}=
l/\kappa_{J_{M}}A$, still considering a system of length $l$ and section $A$.

Let us now consider the opposite configuration, when the animal experiences an
exhausting effort. 
%we can derive the \emph{exhaustion metabolic conductivity}.
In this case, the animal dissipates all the mechanical power while producing
zero contribution to motion. This situations occurs for metabolic intensities
far above the location of the maximal power. The organism is no longer able to
sustain the required effort. The net mechanical power transferred to the environment vanishes, 
so does the net associated mechanical force, i.e. $F_{M}=0$. The entire metabolic power produced by the
muscles is consumed within the animal. In other words, the animal metabolism
is in the short-circuit configuration, fully loaded, and the metabolic
intensity density $J_{M}$ reaches a critical value $J_{X}$ at the exhaustion
stage. Then, from Eq.~(\ref{JMJem2}), the associated energy flux, denoted
$J_{Em}^{\dagger}$, under the zero mechanical condition is:
\begin{equation}
\label{jem3}J_{Em}^{\dagger}=-\kappa_{F_{M}}\frac{\mathrm{d}\mu}{\mathrm{d}x} 
\end{equation}
\noindent where ${\displaystyle
  \kappa_{F_{M}}=\frac{L_{22}}{\mu^{2}}}$ is the exhaustion metabolic conductivity.

One can notice that the link between $\kappa_{F_{M}}$ and $\kappa_{J_{M}}$
derives from the definitions of the transport coefficients:%

\begin{equation}
\label{kappaFM}\kappa_{F_{M}}=\kappa_{J_{M}}\left[  1+\frac{\alpha^{2}\sigma
}{\kappa_{J_{M}}}\mu\right]
\end{equation}

\noindent This relation shows that the most efficient metabolic conditions are when the basal metabolism 
is as low as possible while the exhausting metabolism conditions are delayed as long as
possible. By ``most efficient'', we mean ``with a minimal entropy
production'', hence maximal possible conversion of the food chemical energy
into useful muscular power. The rejected fraction of matter and energy into
the sink is here considered as a waste. It is clear that, contrary to the
waste heat of steam engine, most of this biological waste should be considered
as secondary metabolites for the waste matter, and warming body contribution
for the waste energy. All these secondary contributions could be included in a
more complex thermodynamic network whose generic building block would be the
present Onsager unit cell. A complete analysis of such a complex thermodynamic
network is out of the scope of the present paper.

Equation (\ref{kappaFM}) also indicates that the ratio $\kappa_{F_{M}}/\kappa_{J_{M}}$ 
provides a direct measure of the efficiency of the equivalent
working fluid, as we obtain an expression of the metabolic figure of merit,
which can be viewed as the biological equivalent of the figure of merit known
in the field of thermoelectricity \cite{Goupil2011}:%

\begin{equation}
f_{M}=Z\mu=\frac{\alpha^{2}\sigma}{\kappa_{J_{M}}}\mu\label{Zmu1}%
\end{equation}

\noindent which is the extension of the
expression of Eq.~(\ref{ZPi}) beyond the equilibrium state. In the case of a complete system of length $l$
and section $A$ (see Fig. \ref{Onsager-Cell}b), the figure of merit now reads%

\begin{equation}
f_{M}=\frac{\alpha^{2}R_{E}}{R_{M}}\mu\label{Zmu2}%
\end{equation}

The figure of merit is often defined as the \textit{degree of coupling} of the
fluxes. In their seminal paper, Kedem and Caplan derived the following
expression of the coupling parameter between the two fluxes involved in the
conversion process \cite{Kedem-Caplan1965}:
\begin{equation}
\label{qqq}q=\frac{L_{12}}{\sqrt{L_{11}L_{22}}}=\sqrt{\frac{f_M}{1+f_M}}%
\end{equation}
\noindent which explicitly includes the kinetic coefficients
{$L_{ij}$}. The figure of merit and the coupling factor $q$ are equivalent in
terms of measure of the efficiency of the system: the higher their (absolute)
values, the better the energy conversion system. This can be evidenced by the
derivation of the local maximal efficiency of the conversion process
$\eta_{\mathrm{max}}$:
\[
\eta_{\mathrm{max}} = \left(\frac{1+\sqrt{1-q^{2}}}{q}\right)^{2}
=\frac{\sqrt{1+f_M}-1}{\sqrt{1+f_M}+1} 
\]
We observe that the figure of merit $f_M$ provides a direct quantitative
indicator of the performance of the considered system. Denoting $F_{M_{0}}$
the macroscopic force under the basal metabolism configuration, $J_{M}=0$, the
expression of the coefficient $\alpha$ may be rewritten
as:
\begin{equation}
\alpha=F_{M_{0}}\left(  \frac{d\mu}{dx}\right)  ^{-1}=\frac{1}{\mu}%
\frac{L_{12}}{L_{11}} \label{alpha2}%
\end{equation}
\noindent From the derivation of the transport coefficients, we see that only
three transport parameters are required to characterize the metabolic machine:
two conductivities, respectively for the aggregated energy flux, $\sigma$, and
the dispersed energy flux, $\kappa_{J_{M}}$, and a coupling parameter between
these two fluxes, namely $\alpha$.

\section{The thermodynamics of metabolism}

\label{sec3}

\subsection{The metabolic machine}

\subsubsection{Local energy budget}

Replacing the kinetic coefficients $L_{ij}$ in Eq.~(\ref{LMJem}) by the
transport coefficients, $\sigma$, $\kappa_{J_{M}}$ and $\alpha$, yields the
following expressions for the fluxes $J_{M}$ and $J_{Em}$:
\begin{equation}
\label{JmJem5b}\left(
\begin{array}
[c]{c}%
J_{M}\\
J_{Em}%
\end{array}
\right)  =\left(
\begin{array}
[c]{cc}%
\sigma & \alpha\sigma\\
\alpha\sigma\mu & \kappa_{\Pi_{M}}%
\end{array}
\right)  \left(
\begin{array}
[c]{c}%
F_{M}\\
-\frac{\displaystyle \mathrm{d}\mu}{\displaystyle \mathrm{d}x}%
\end{array}
\right)
\end{equation}

\noindent Assuming constant parameters,
%\bibitem{noteCPM}
the gradient of $J_{Em}$ then reads
\begin{equation}
\label{dJem1}\frac{\mathrm{d}J_{Em}}{\mathrm{d}x}=\alpha J_{M}\frac
{\mathrm{d}\mu}{\mathrm{d}x}-\kappa_{J_{M}}\frac{\mathrm{d}^{2}\mu}%
{\mathrm{d}x^{2}}%
\end{equation}
\noindent Note that the transport coefficients are in general not constant, but the
constant transport parameter assumption is routinely used when the potential
gradients in a given system are sufficiently small so that the system remains
close to equilibrium (or close to some effective quasi-equilibrium). For sake of simplicity the
present work focusses on ``moderate'' efforts, which imply small gradients of the intensive parameters
and hence a negligible dependence of the transport coefficients on energy,
space and time, as one analyses the production of mechanical force on the
global scale. Also, we make the assumption of sufficiently short duration
metabolic effort and hence secondary metabolites do not partake in the
chemical-to-mechanical energy conversion and are simply considered as waste.
In this case, the transport parameters are not affected by the presence of
these secondary metabolites.

\noindent As energy conservation imposes a constant total energy flux $J_{U}$
in Eq.(\ref{JU1}), we get%

\begin{equation}
\label{dJem2}\frac{\mathrm{d}J_{Em}}{\mathrm{d}x}=F_{M}J_{M}%
\end{equation}

\noindent and combining Eqs. (\ref{dJem1}) and (\ref{dJem2}), we finally
obtain the local energy budget:%

\begin{equation}
\label{JM4}\kappa_{J_{M}}\frac{\mathrm{d}^{2}\mu}{\mathrm{d}x^{2}}%
=-\frac{J_{M}^{2}}{\sigma}%
\end{equation}

One may now notice that although dissipation does not explicitly appear in the
force-flux expressions, it does in the local budget equation through the term
$J_{M}^{2}/\sigma$. In other words, the conservation of the energy and matter,
$\nabla{J_{U}}=0$ and $\nabla{J_{M}}=0$ and the local working conditions of
the metabolic device are totally defined by the local budget and the three
transport coefficients $\sigma$, $\kappa_{J_{M}}$ and $\alpha$. Let us now
extend the analysis to the management of this equivalent working fluid inside
an organism or part of it.

\subsubsection{Metabolic flux}

Assuming that the metabolic processes take place in the volume $A\times l$ of
the organism, the complete energy budget is obtained after integration of
Eq.~(\ref{JM4})%

\begin{equation}
\label{dmu}\frac{\mathrm{d}\mu}{\mathrm{d}x}=-\frac{J_{M}^{2}}{\sigma
\kappa_{J_{M}}}x+C
\end{equation}

\noindent The energy flux $\Phi=AJ_{Em}$ can hence be expressed as%

\begin{equation}
\label{Phi}\Phi=\alpha\mu I_{M}-\kappa_{J_{M}}A\frac{\mathrm{d}\mu}%
{\mathrm{d}x}%
\end{equation}

\noindent where $I_{M}=AJ_{M}$ is the metabolic intensity. This latter is directly related to the average rate of production of the chemical reactions. The metabolic flux
now reads from Eq.~(\ref{dJem2}):%

\begin{equation}
\label{Phix}\Phi(x)=\alpha\mu(x)I_{M}+R_{M}I_{M}^{2}\frac{x}{l}+C
\end{equation}

\noindent The term $C$ is determined by the boundary conditions at both
locations $x=0$ and $x=l$. Now, in order to complete the description of the
system and the model, we turn to boundary conditions specification.

\subsection{The metabolic system}

\subsubsection{Metabolic power}

\begin{figure}[ptbh]
\centering
\includegraphics[width=0.7\textwidth]{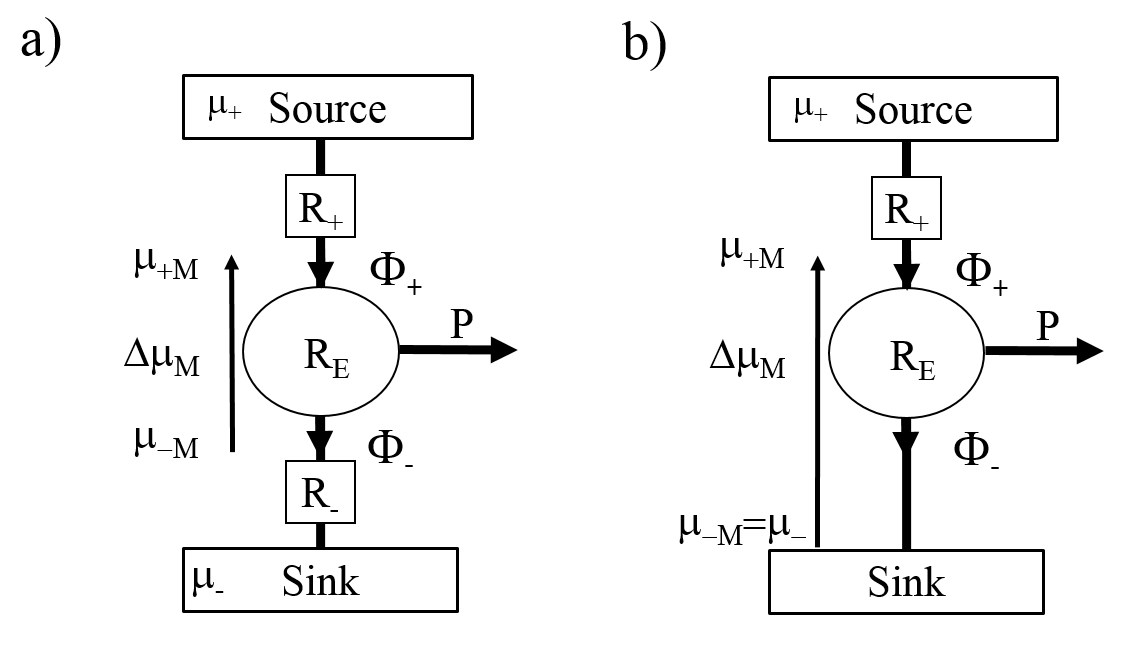}\caption{Thermodynamic
system: a) general configuration. b) simplified configuration for
low-duration, steady-state efforts, with $R_{-}=0$. A dissipative coupling modifies the
chemical potential difference across the conversion zone.} 
\label{YCNCA} 
\end{figure}

As a thermodynamic device, the metabolic conversion zone defined by the
element of length $l$ and section $A$ is inserted in a global system. The
boundary conditions impose the values of both energy and matter currents and
consequently, the working conditions of the metabolic device. The complete
thermodynamic system is represented on Fig.~\ref{YCNCA}, which
arguably % indeed
outlines a quite general situation. Due to the presence of dissipative
couplings between the conversion zone and the reservoirs, the reservoirs
chemical potentials $\mu_{+}$ and $\mu_{-}$ are modified and become $\mu_{+M}$
and $\mu_{-M}$ as shown on Fig.~\ref{YCNCA}.

The metabolic converter is connected to a reservoir and a sink by two
dissipative elements of resistances, $R_{+}$ and $R_{-}$. These resistances,
together with the basal metabolic resistance $R_{E} = l/\kappa_{J_{M}}A$,
allow considering mixed boundary conditions between two limit cases, namely i)
the Neumann conditions, where currents are imposed, considering high,
diverging values of $R_{+}$ and $R_{-}$; ii) the Dirichlet conditions, where
potentials are imposed, considering instead vanishing values for $R_{+}$ and
$R_{-}$. It is important to note that the crucial question of boundary
conditions is imposed by the system, namely, a metabolic machine contained
within a defined perimeter. From a theoretical point of view, it would be
perfectly possible to define conditions at Neuman's limits by considering the
organism within its biotope, as a complete system receiving a flow of matter
and energy. However, the perimeter of the system, imposed by the body envelope
is an unavoidable constraint in this model, and this constraint leads to
consider irreducible mixed boundary conditions. From the general expression
Eq.~(\ref{Phix}), we get the expressions of the incoming $\Phi_{+}=\Phi(0)$
and outgoing $\Phi_{-}=\Phi(l)$ fluxes:%

\begin{eqnarray}
\label{Phiplus}\Phi_{+}  &  = \alpha\mu_{+M}I_{M}+\frac{\Delta\mu_{M}}{R_{E}%
}\\
\label{Phimoins}\Phi_{-}  &  = \alpha\mu_{-M}I_{M}+R_{M}I_{M}^{2}+\frac{\Delta\mu_{M}}{R_{E}}%
\end{eqnarray}

\noindent where $\Delta\mu_{M}=\mu_{+M}-\mu_{-M}$. \noindent The term
$C=\Delta\mu_{M}/R_{E}$ comes from the flow balance at the incoming and
outgoing borders of the conversion zone. Accounting for the connections to the
reservoirs, we get two additional expressions for the fluxes:%

\begin{eqnarray}
\label{Phi12}\Phi_{+}  &  = \frac{\mu_{+}-\mu_{+M}}{R_{+}}\\
\label{Phi13}\Phi_{-}  &  = \frac{\mu_{-M}-\mu_{-}}{R_{-}}
\end{eqnarray}

\noindent that yield the following simple expression for the metabolic
converter output power $P$:%

\begin{equation}
\label{Pout}P=\Phi_{+}-\Phi_{-}=\left[  \alpha\Delta\mu_{M}-R_{M}I_{M}\right]
I_{M}%
\end{equation}

\noindent The expressions (\ref{Phiplus}), (\ref{Phimoins}), (\ref{Phi12}), 
and (\ref{Phi13}) now give a complete set of equations for a proper description
and analysis of a given metabolic situation. The calculation of the general
solution does not pose any specific difficulty but is quite tedious.

\subsubsection{Chemical-to-mechanical energy conversion parameters}

For illustration purposes, let us consider the particular situation of an
effort with reasonable (low) duration, neglecting the drawback effects of the
waste fractions; in this case, the coupling resistance to the sink $R_{-}$ may
be considered as negligible, and $\mu_{-M}\approx\mu_{-}$. We obtain the
simplified expressions:%

\begin{eqnarray}
\label{FLUXES1}\Phi_{+}  &  = \frac{F_{\mathrm{iso}}I_{M}/{\eta}_{\mathrm{C}%
}+B}{I_{T}+I_{M}}~I_{T}\\
\label{FLUXES2}\Phi_{-}  &  = R_{M}I_{M}^{2}+\frac{R_{\mathrm{fb}}I_{M}^{2}+{F_{\mathrm{iso}%
}}\left(  \frac{1}{{\eta}_{\mathrm{C}}}-1\right)  I_{M}+B}{I_{T}+I_{M}}~I_{T}%
\end{eqnarray}

\noindent where all the quantities are defined in Table \ref{table2}. In order to simplify the identification 
of the boundary conditions we define the
parameter $r=\frac{R_{+}}{R_{+}+R_{E}}$. This parameter varies from $r=0$ when
the system is connected with potential boundary conditions, i.e. of the Dirichlet type, 
to $r=1$ in the case of flux conditions, i.e., Neumann type.

\begin{table}[th]
\centering{\scriptsize
\begin{tabular}
[c]{ll}\hline\hline
Quantity~ & ~ Definition\\\hline
~ & ~\\
${\displaystyle I_{M}}$ ~ & ~ macroscopic (space integrated) metabolic
intensity\\
~ & ~\\
${\displaystyle I_{M}F_{M}}$ ~ & ~ available mechanical power\\
~ & ~\\
${\displaystyle B=\frac{\Delta\mu}{R_{+}+R_{E}}=}\frac{\Delta\mu}{R_{E}}(1-r)$
~ & ~ basal power consumed by the body at $I_M=0$, or $F=F_{\rm iso}$\\
~ & ~\\
${\displaystyle{I_{T}=\frac{R_{E}+R_{+}}{\alpha R_{+}R_{E}}=}}\frac{1}{\alpha
rR_{E}}$ ~ & ~ threshold metabolic intensity\\
~ & ~\\
${\displaystyle R_{\mathrm{{fb}}}=\alpha\mu_{-}/I_{T}=r}\alpha^{2}R_{E}\mu
_{-}$ ~ & ~ feedback resistance\\
~ & ~\\
${\displaystyle F_{\mathrm{iso}}{=\alpha R_{E}B}}$ =$\alpha{\Delta\mu}%
(1-r)$~ & ~ isometric force\\
~ & ~\\
${\displaystyle\eta_{\mathrm{C}}=\frac{\mu_{+}-\mu_{-}}{\mu_{+}}}$ ~ & ~
overall chemical-mechanical machine efficiency\\
~ & ~\\\hline\hline
\end{tabular}
}\caption{Thermodynamic parameters characterizing the chemical-to-mechanical
energy conversion.}%
\label{table2}%
\end{table}

\noindent Note that the product or ratio of some couples of parameters result in constant
quantities such as, e.g., $F_{\rm iso}/B=\alpha R_{E}$ and $R_{\rm fb}I_{T}=\alpha\mu_{-}$, 
which creates some particular constraints. This point will be developed further in the article.

Both $R_{\mathrm{fb}}$ and $I_{T}$ are governed by the boundary conditions for
the energy entering the system; in other words, they characterize the ability
of the biological system to feed the conversion zone with chemical energy,
this ability being limited by the coupling resistance $R_{+}$. Note that if
the boundary conditions would only be of the potential type (Dirichlet
conditions), the feedback resistance
$R_{\mathrm{fb}}$ would be zero. We discuss the importance of the feedback
resistance for muscle work in the frame of Hill's model in the next section,
as $I_{T}$ corresponds exactly to one of Hill's constants, and $R_{\mathrm{fb}%
}$ is instrumental to elucidate the question of the non-constancy of
Hill's first \textquotedblleft constant\textquotedblright\ $a$ (defined in the
next Section).

The metabolic power delivered during a physical effort is expressed as:%

\begin{equation}
\label{PetFLUX}P=\Phi_{+}-\Phi_{-}=F_{M}I_{M}=\left[  {F_{\mathrm{iso}}%
}-\left(  R_{M}+R_{H}(I_{M})\right)  I_{M}\right]  I_{M}%
\end{equation}

%The expression of the efficiency is,
%\begin{equation}
%\label{Efficiency}\eta=\frac{\Phi_{+}-\Phi_{-}}{\Phi_{+}}=???????????????????????
%\end{equation}

\noindent where $F_{\mathrm{iso}}$ is the so-called isometric force
\cite{Curtin2015}, which describes the situation of a muscle tension under
load but with no motion, and where the resistance $R_{H}(I_{M})=\frac
{{F_{\mathrm{iso}}}+R_{\mathrm{fb}}{{I_{T}}}}{I_{T}+I_{M}}$ is introduced as
 a metabolic intensity-dependent resistance showing clearly that
the organism cannot be seen as a passive system in the sense that it does not
merely represent a system component that merely dissipates energy as a
standard resistor would in an electrical circuit. In the present model,
$R_{M}$ acts as such a passive component. On the contrary, the definition of $R_{H}$
contains $R_{\mathrm{fb}}$, and both can be expected to stem from feedback
effects. They are thus tightly related, notably through the threshold
metabolic intensity.

With Eq.~(\ref{PetFLUX}), we also recover the classical expression of a
force-intensity response, with the isometric force ${F_{\mathrm{iso}}}$ and
the composite internal resistance $R_{H}(I_{M})+R_{M}$. As expected, the
isometric force is proportional to the converted fraction of the chemical
energy difference $\Delta\mu$ modulated by the resistance bridge value:
$\frac{R_{E}}{R_{+}+R_{E}}$. Furthermore, as the composite resistance $R_{H}(I_{M})+R_{M}$ depends
on the metabolic intensity, we see that it may govern the shape of the response
curves of a muscle, which is quite a central result, as will be shown later.

The output power accounts for the power spent to sustain the physical effort
balanced by the dissipated power. It is important to note that the notion of
``physical effort'' may encompass a quite large variety of efforts. In the
case of an animal in motion, at velocity $v$, the physical effort simply corresponds to the
production of the mechanical power necessary for actual motion. In this case,
the metabolic intensity can be expressed directly from the rate of production of chemicals \cite{Barany1967}
and, consequently, the mechanical velocity of the system as $I_{M}\equiv kv$, where $k$ is a dimensional constant. But the
metabolic intensity is also non-zero for animals carrying heavy loads, while
standing still. Further, the power $P$ has two zeros, corresponding to two
specific values for the intensity: $I_{M}=0$ in the absence of any effort, and
$I_{M}=I_{X}$, the maximum value of intensity at $P=0$, in a situation of
exhaustion with $I_{X}=\frac{{F_{\mathrm{iso}}}}{R_{H}+R_{M}}$. Therefore, the
concept of metabolic intensity accounts for various types of efforts, that a
purely mechanical description cannot.

Turning to the figure of merit, we see that it now reads
\begin{equation}
f_{M}=\frac{R_{\mathrm{fb}}F_{\mathrm{iso}}I_{T}}{R_{M}B}
\label{RfbZmu}%
\end{equation}
\noindent
We thus obtain a compact expression which, while preserving its thermodynamic
structure, can be read in a form directly accessible to the understanding of
metabolic performance. Let us analyse what are the compromises and expectations at stake
for a given value of this merit
factor.
%for. % the resulting expectations.
 First of all, it is clear that mechanical
power is only available as much as the mechanical viscosity $R_{M}$ makes it possible.
Similarly, we can notice the presence of the $F_{\mathrm{iso}}/B$ ratio which
reflects the necessary compromise between the availability of force
and the ``basal'' 
dimension of the machine that leads to this availability. Last, the product
$R_{\mathrm{fb}}I_{T}=\alpha\mu_{-}$ which can be considered constant, shows us
that the metabolic intensity is necessarily contingent on the feedback
resistance, and that high metabolic intensities can only be achieved by a
significant reduction in this resistance.

\begin{figure}[ptbh]
\begin{center}
\includegraphics[width=0.8\textwidth]{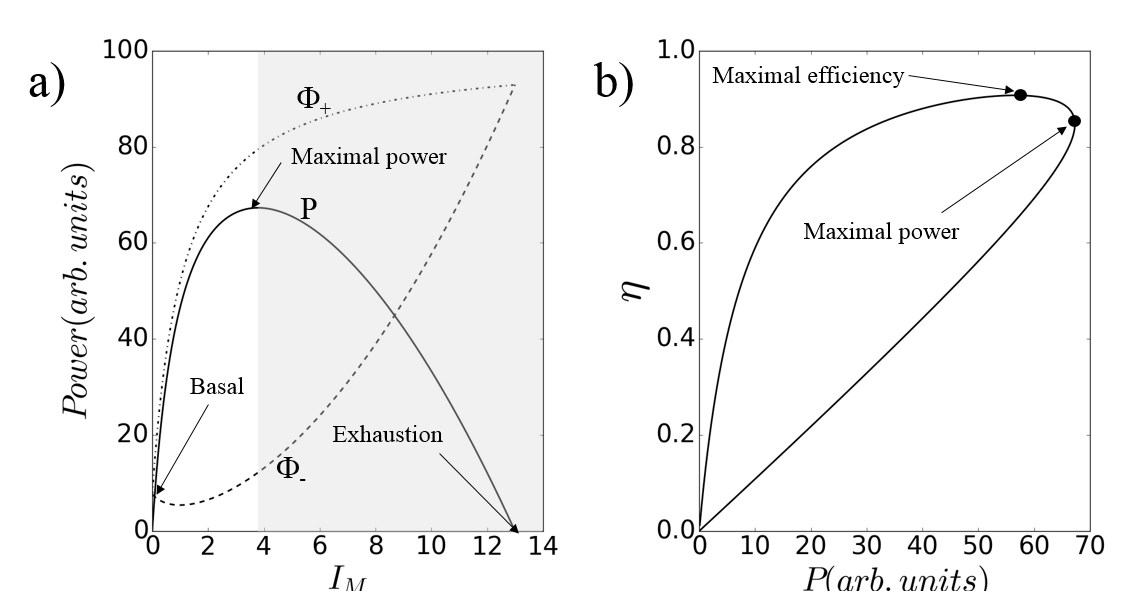}
\end{center}
\caption{a) Generic power plots: external mechanical power $P$ (straight
line), input $\Phi_{+}$ (dash-dotted), and output $\Phi_{-}$ (dash) metabolic
fluxes. The shaded area corresponds to the region where the organism degrades
more mechanical power than it uses for the effort. b) Efficiency versus power
response. As the metabolic intensity $I_{M}$ increases, the efficiency $\eta$
reaches its maximum before the mechanical power reaches its own maximum.}%
\label{courbes}%
\end{figure}

It is important to note that $f_{M}$ does not depend on the ratio $r= R_E/(R_+ + R_E)$; in fact, 
it is a quantity that intrinsically characterizes the thermodynamic conversion
capacity, regardless of the coupling conditions to the reservoirs. Finally,
using the set of equations (\ref{FLUXES1}), (\ref{FLUXES2}) and (\ref{PetFLUX}), 
we can now plot the generic response of an organism to an effort as shown
on Fig. \ref{courbes}a where the input ($\Phi_{+}$), output ($\Phi_{-}$) and
mechanical ($P=\Phi_{+}-\Phi_{-}$) powers are reported. We also plot on
Fig. \ref{courbes}b the efficiency $\eta=\frac{\Phi_{+}-\Phi_{-}}{\Phi_{+}}$ \emph{vs} power 
curve: the maximal efficiency is obtained for a metabolic intensity 
%slightly smaller than  
substantially  below that
required for the production of maximal mechanical power, 
while the mechanical output is only slightly below the maximal one.
The latter is reached
at the cost of efficiency. It can easily be shown that the higher the $f_M$
factor, the further away the maximum efficiency and maximum power
points shall be apart. This
confirms that there is no optimal operating point in absolute terms, let alone
an underlying general variational principle. Depending on the metabolic
parameters, each muscle is subjected to a trade-off between its maximal
efficiency and its maximal power, or minimum production of waste. We shall
discuss the question of the variational principle in Section
\ref{sec5}. We show that the biological system becomes obviously deterministic at the condition 
that the system definition properly includes the boundary conditions.

\section{Metabolic energy conversion under muscle load}

\label{sec4}

\subsection{Physical efforts}

\begin{figure}[ptbh]
\centering
\includegraphics[width=0.7\textwidth]{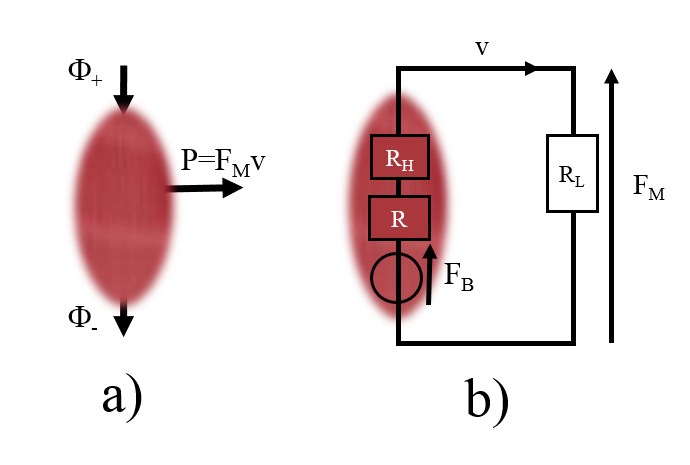}\caption{Thermodynamic
to mechanical conversion: a) Metabolic power conversion; b) Mechanical power
production.}%
\label{Global}%
\end{figure}

Let us now consider the muscular strength deduced from the expression of power
by simply writing $F_{M}=P/I_{M}$. As previously mentioned, using the above formalism, we can identify
the force from an equivalent generator scheme, including in series the isometric
force $F_{\mathrm{iso}}$ and the internal resistance $R_{M}+R_{H}(I_{M})$. It
follows that Figure \ref{Global} summarizes the principles of the
thermodynamic model of metabolic power generation, i.e. energy
conversion from
chemical power to mechanical power, in the case of short-duration efforts. The complete system is composed of two
parts. The first one (a) is the biochemical converter, which receives the incoming
chemical power $\Phi_{+}$, delivers an output power $P$, and produces waste
and secondary metabolites, denoted $\Phi_{-}$. In b), the
mechanical power $P$ is partly dissipated within the organism because of
``friction'' due to the internal composite resistance
$R_{M}+R_{H}(I_{M})$, where the latter contribution corresponds to the
variable part of the dissipative mechanism. Without going into details that
are beyond the scope of this article, we observe that this non-linear mechanical
resistance is an \emph{active} impedance in the sense that it depends on the metabolic intensity, and hence on the
operating point of the system. This is 
a major and principle departure from
%in contradiction with 
passive models
that traditionally %represent
take this restrictive assumption on
 this impedance. 
Moreover, several experimental
studies based on the frequency-dependent linear response of this impedance
\cite{Cornu1997,Goubel1998,Desplantez1999,Desplantez2002}, show a resonant
frequency response. In the presence of a feedback, such a resonance is quite possible
since the response of the closed-loop system may present an even higher order
response than that of an open loop system \cite{Goupil2016}. 
At this stage of the description, it becomes important to compare the model and its predictions 
with a well-defined biological system. To do so, we consider the 
case of skeletal muscle in the frame of Hill's approach.

\subsection{Revisiting Hill's muscle load model} 

Assuming that the entire metabolic power produced is converted into mechanical
power, we may write $P=F_{M}I_{M}$, and using Eq.~(\ref{PetFLUX}), we find:

\begin{equation}
F_{M}=\frac{\left(  F_{\mathrm{iso}}+R_{\mathrm{fb}}I_{T}\right)  I_{T}%
}{\left(  I_{M}+I_{T}\right)  }-\left(  R_{\mathrm{fb}}I_{T}+R_{M}%
I_{M}\right)  \label{FM}%
\end{equation}

\noindent This expression is the thermodynamics-based formulation of the force response
in presence of an effort of intensity $I_{M}$. We now turn to the comparison
with the model proposed by Hill in 1938 \cite{Hill1938}, which is the most
classical muscle description for chemical-to-mechanical power metabolic
conversion, and thus serves here as a touchstone. In his seminal article based on dynamic
calorimetric measurements, Hill assumed that the chemical energy rate was proportional
to the contraction velocity $v$ of the muscle, and consequently that: $I_{M}\propto v$. In
other words he proposed to switch from thermodynamic arguments to mechanical
ones (see \ref{annexe2}). Hill's state equation reads:

\begin{equation}
F_{M}=\frac{c}{\left(  v+b\right)  }-a\noindent\label{FHILL}%
\end{equation}
From Eq. \ref{FM} we unambiguously identify
\begin{equation}%
a =R_{\mathrm{fb}}I_{T}+R_{M}I_{M}\noindent\label{a}%
\end{equation}%
\begin{equation}%
b =I_{T}\noindent\label{b}%
\end{equation}%
\begin{equation}%
c = \left(  F_{\mathrm{iso}}+R_{\mathrm{fb}}I_{T}\right)  I_{T}\noindent\label{c}% 
\end{equation}

Besides the similarity of Eq.~(\ref{FM}) and Eq.~(\ref{FHILL}) it is
important to consider how to give Hill's model a proper thermodynamic ground. 
In this spirit, we propose to consider four
questions that mark the history of the Hill model and that turn out to find a common
framework here: i) the chemical origin of part of the observed mechanical
dissipation; ii) the presence of slow and fast muscle fibres, iii) the rectilinear versus non-linear
shape of the force-velocity measurements, and, finally iv) the
speed dependence of the so-called 
\emph{extra heat} term $a$.

\subsubsection{Chemical origin of parameter $a$}
The principal elements of Hill's model are summarized in \ref{annexe2}; the model shows that a
chemical contribution to dissipation during muscle contraction was a central
hypothesis of the muscle model \cite{Hill1938}, and as such, had been added to
the energy budget. This point had first been raised by Fenn who claimed 
that the fact that an active muscle shortens more slowly under a greater force,
is not due to mechanical viscosity but to how the chemical energy release is regulated \cite{Fenn1923,Fenn1935} 
(see also the critical analysis of Fenn's works in \cite{Rall1982}). 
Both Fenn and Hill did not make use of any feedback effect
between chemical input and mechanical output. However, considering the expressions of 
$R_{\rm fb}$ and $I_T$, the chemical origin of $a$ is evident.
In 1966, Caplan conducted a similar analysis, but the work done considered the 
local feedback to be an exogenous additional process, 
%with no account for any change
without any hint to a change
in boundary conditions applied to a macroscopic size system \cite{Caplan1966}. 
The influence of modified chemical conditions can also be found in \cite{Fitts1991} and \cite{Alexander1977}.

\subsubsection{Fast and slow fibers} 
In his Nobel lecture Hill put forth the central question of the fast
and slow muscle fibers \cite{HillNobel}:\\

``\emph{The difference in the time-scales of the two types of muscle makes one regard it as improbable that physical 
viscosity alone is the determining factor. One cannot see why viscosity should have ten times the effect in a human muscle 
than it has in a frog's, and probably one hundred or one thousand times as much as it has in a fly's.}''\\

From a mechanical point of view there is no reason why a
machine should be constrained by limitations as long as it is able to receive
and convert energy, such as chemical energy, into mechanical
energy. On the other hand, from the point of view of
thermodynamics, the situation is quite different. Indeed, since the
performance of the conversion machine is defined both by the thermodynamic
fluid and by the machine that uses this fluid, it follows that the maximum
performance is limited by the intrinsic capacity of the fluid to carry out the
energy conversion. The constant figure of merit $f_{M}=\frac{R_{\mathrm{fb}}F_{\mathrm{iso}}I_{T}}{R_{M}B}$ 
reflects these intrinsic performances. For a general thermodynamic system there is a figure of merit  
that defines the upper limit of the performance of the conversion to
useful work. 
This limit can be approached but never exceeded. The situation here is strictly similar and there is 
therefore a trade-off as to the values of the
different parameters. The performance of a muscle is then bounded from above
by the figure of merit, that defines the maximum thermodynamic
conversion capacity, and consequently, the maximum intrinsic efficiency. To
this must be added the constraints that link some of the parameters that make
up the figure of merit, namely: $F_{\rm iso}/B=\alpha R_{E}$ and
$R_{\rm fb}I_{T}=\alpha\mu_{-}$, which are considered constant.
 Indeed, in the present case of the response to a time-limited effort, 
the intrinsic parameters $R_E$, $\mu_-$ and $\alpha$ are constant. 

Let us now consider the specific case of a so-called fast muscle and observe the constraints 
imposed by the metabolic model. In this case it is expected that the force does not collapse at 
high speed $v$, i.e. for high values of metabolic intensity close to $I_{T}$. In view of the above remarks, 
this implies that $R_{\rm fb}$ is minimal, and then so is $R_{+}$. This leads to an increase
in the basal power $B$ and therefore an increase in the isometric force $F_{\rm iso}$
by the same amount. Therefore, it appears that fast fibres are also those with
the highest isometric forces, and as such are the
most powerful fibers, which benefit from little-constrained access to the
resource since $r\ll 1$. This double signature is indeed encountered in the case of fast fibres, 
which are highly energy consuming, even at low speeds, and in particular at the basal level \cite{Close1964,Close1972}. 
Then the performance is achieved at the cost of maintaining 
a substantial basal power that has a definite energy cost which in
turn reduces the overall efficiency. 
On the contrary, in the case of slow fibres, the stress on large $I_{T}$ values is released which
leads to a higher $R_{\rm fb}$ resistance and finally a moderate basal power. Note that at birth, 
some mammals have a very high proportion of slow muscles \cite{Close1964}, which tends to decrease as they grow. 
Such a signature at birth is in accordance with the main property of slow muscles which is their low basal consumption. 
The same conclusions can be found in the works of Ruegg \cite{Ruegg1971} and Clinch \cite{Clinch1968}. 

The difference in the respective values of the basal consumption of fibres and their contraction rates is commonly 
reported in the literature. As an example we cite the case of the muscles ``extensor digitorum longus'' and ``soleus'', 
which in mice have contraction rate ratios of 5.9/2 $\simeq$ 2.95  and
basal consumption in the ratio 4/1.3 $\simeq$ 3.08 (a value quite
close to 2.95) \cite{Kushmerick1983,Crow1982,Crow1983}. Since the basal power  %, which corresponds to an engine in idle state, 
does not contribute to the production of mechanical power, the choice of slow fibres, when they are sufficient to achieve the function, 
becomes obvious. %just as obvious as that of a small displacement engine for a slow vehicle. 
From the point of view of glucose combustion, the glycolitic pathway is incomplete since it does not include a Krebbs cycle. 
The similarity with complete or incomplete combustion of alkanes in internal combustion engines, is quite natural, the efficiency 
ceiling being in both cases set by the figure of merit of the fuel under total combustion conditions. In other words, one may observe 
that the overall performance of the muscle is, as in the case of any thermodynamic system, the result of a trade-off between the intrinsic 
combustion properties of the fuel on the one hand, and the implementation of this combustion according to a particular path imposed by the 
machine, on the other hand. Similarly, we can see that in the case of muscles of the same nature, ($r$ constant) the ratio $a_{0}/F_{\rm iso}$ 
is predicted to be constant, which is actually reported in \cite{Close1964}. Although temperature is not explicitly present in our model, 
the influence of a muscle temperature during exercise is easily visible on the measurements \cite{Ranatunga2018}. The general shape 
of the curves is not globally modified, but the parameters have a
sizable temperature dependence. 

\subsubsection{Linear and non-linear shape of the force-velocity curves}
It is known that slow and fast fibres do not have the same force-speed
response. If the classic form of the force-speed response according to Hill is
that of a hyperbola, mention has been made of responses with a very shallow,
or even totally rectilinear, character \cite{Fitts1991,Close1964,Close1972}. The existence of linear
characteristics means that in these cases the dissipation resistance
$R_{H}(I_{M})+R_{M}$ is almost constant. As a result, the term $R_{H}(I_{M})$
is no longer a function of the current $I_{M}$. This amounts to considering
that $I_{T}\gg I_{M}$. It follows that the predominance of
fast fibres leads to more linear characteristics. This was indeed observed for
the human in the force-speed response of the arms \cite{Sreckovic2015}. Physiological analyses
reveal the predominance of fast fibres in the arms, unlike legs which have more slow fibres.
This is true in general for the untrained individual. 
In the case of sprint-trained runners \cite{Samozino2016} or cyclists \cite{Dorel2005},
 there is a strong tendency towards the linear form, which in these cases reflects the predominance of fast fibres, 
as predicted by the model. The same signature is obtained if the pedalling is done with the arms \cite{Vandewalle1987}.

We can illustrate the dependence of the boundary conditions connection quality (value of $r$) according to the fibre 
types by using data on EDL and soleus muscles \cite{Close1964}. These latter are distinguished
by their fraction of fast and slow fibres, which in our formalism translates as a better coupling to the reservoirs, 
and hence: $r_{\rm{SOL}} < r_{\rm{EDL}}$. Further, by normalizing the force by the isometric force, $F_{\rm{iso}}$, 
in Eq.~(\ref{FM}), and the velocity by the short circuit velocity, $v_X$, we obtain the following expression:
\begin{eqnarray}
  \frac{F}{F_{\mathrm{iso}}} & = & \zeta (r)  \frac{1 - v / v_X}{v / v_X + \zeta
  (r)} - \frac{R_M v_X}{F_{\mathrm{iso}}}  \frac{v}{v_X} 
  \label{eq:hillad}
\end{eqnarray}

\noindent with $\zeta (r) = \frac{a_0}{F_{\mathrm{iso}}} = \frac{a_0}{\alpha \Delta \mu} 
\frac{1}{1 - r}$ and $\frac{a_0}{\alpha \Delta \mu}$ constant. The first term
corresponds to the force-velocity response in the absence of contribution of the mechanical dissipation, i.e., $R_M =0$. 
We then obtain an expression whose only degree of freedom is $r$. In other words, one can only distinguish a fast fibre 
from a slow fibre, or two muscles composed of a different fraction of each other, by their dependence on boundary conditions.
The presence of mechanical dissipation ($R_M \neq 0$) results in a correction for the higher velocity: for a given velocity, 
the available force is reduced. 

\begin{figure}[ptbh]
\centering
\includegraphics[height=0.7\textwidth]{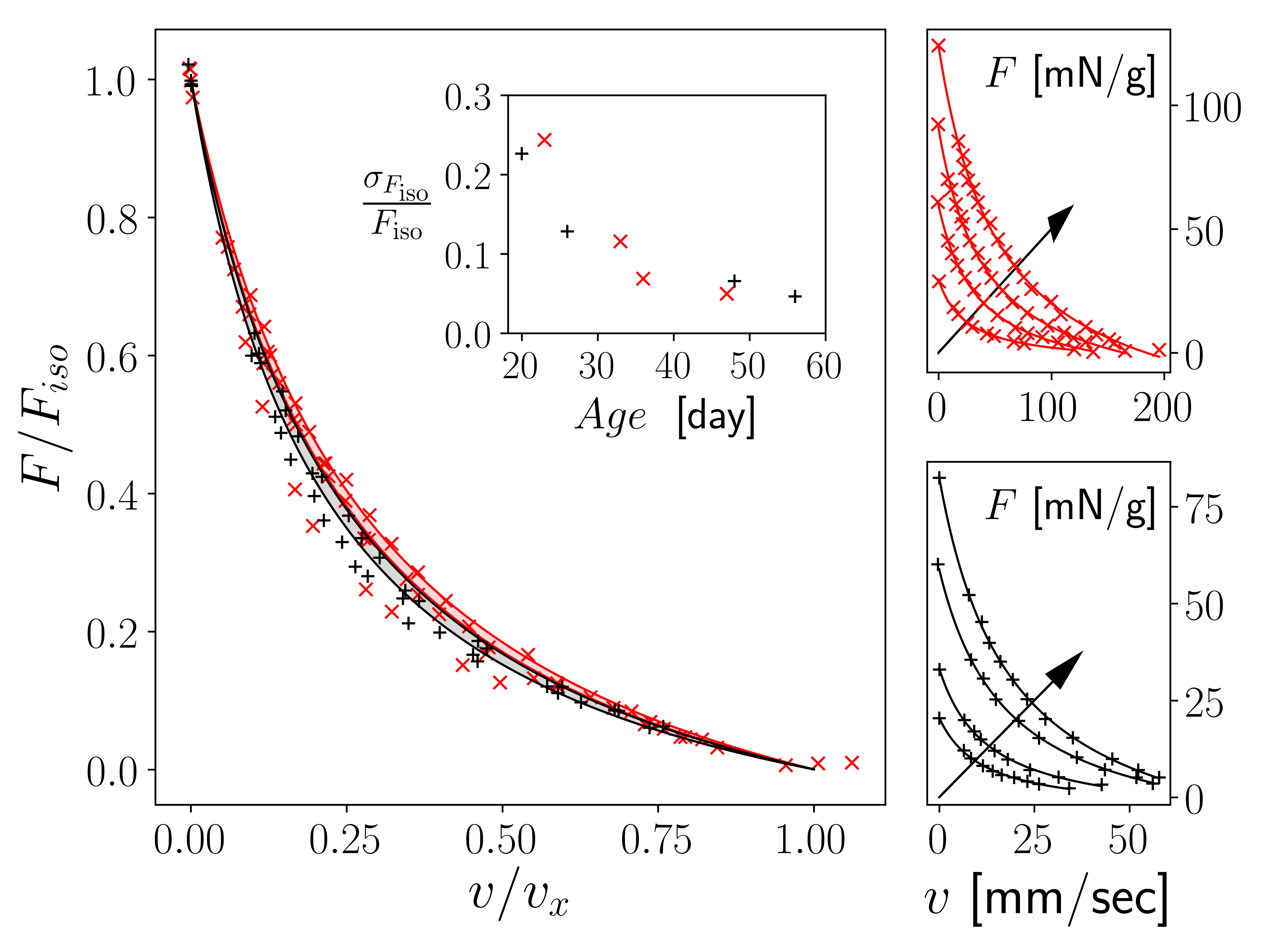} \\
\caption{On the right panel, we show the data extracted from \cite{Close1964}. Top 
is force-velocity response for the \textit{extensor digitorum longus}
(EDL, red $\times$) and bottom is \textit{soleus} (SOL,  dark $+$) of
female rats for various ages (resp. 23, 33, 36 and 47 days, and 20,
26, 48 and 56 days). 
In both cases, age increases in the direction of the arrow. 
Lines are respective best fits, from Eq.~(\ref{FHILL}) without
dissipation ($R_M=0$ or $a=a_0$). 
The corresponding relative uncertainty of $F_{\rm iso}$ compared to
the age of the muscle is represented  on the inset of the left panel. 
On the left panel is shown the corresponding normalized force-velocity
response for EDL and SOL using the same color and symbol codes. 
Normalization is obtained by extracting for each set of data the
respective force $F_{\rm iso}$ and velocity $v_x$, which is the maximum
velocity at $P=0$. 
The respective shaded areas correspond to curve fitting, 
based on Eq.~(\ref{eq:hillad}), which also takes into account
uncertainties (see inset) in data extraction 
and curve fitting of the force-velocity response.}
%Right panel,
%% adapted 
%data extracted from \cite{Close1964}: 
%Bullets are force-velocity response for the extensor digitorum longus (EDL, top, $\times$ red) and soleus (SOL, bottom, $+$ black) of female rats for various ages (resp. 23, 33, 36 and 47 days, and 20, 26, 48 and 56 days). In both cases, the isometric force $F_{\rm iso}$ increases with time. Lines are respective best fits, from Eq.~(\ref{FHILL}) without dissipation ($R_M=0$ or $a=a_0$). Left panel: Corresponding normalised force-velocity response for EDL (in black) and SOL (in red). The normalisation is obtained by extracting for each set of data the respective force $F_{\rm iso}$ and velocity $v_X$, which is the the maximum velocity at $P=0$. The respective shaded areas correspond to curve fitting, based on Eq.~(\ref{eq:hillad}), which take into account uncertainties (see inset) in data extraction and curve fitting of the force-velocity response.}
\label{fig:Close}
\end{figure}

The panels on the right side of Fig.~\ref{fig:Close} show experimental data on the EDL and soleus muscles \cite{Close1964}, 
in a force-velocity representation. The fits shown are based on Hill's equation without direct contribution of the mechanical 
dissipation, for which $R_M=0$ and $a = a_0$; this hypothesis makes it possible to fit the experimental data with very good agreement. 
In the following, we limit the scope of our analysis to this configuration. 

The experimental data normalized by the isometric force, $F_{\mathrm{iso}}$, and the short circuit velocity, $v_X$, obtained for each of
the previous fit are reported on the panel left of Fig.~\ref{fig:Close}. The fits of these data are also done with Eq. \ref{eq:hillad} under 
the assumption of mechanical dissipation. The uncertainty of the experimental data was estimated at 1/4 of the marker size. 
We observe in the inset of Fig.\,\ref{fig:Close} that both relative and absolute uncertainties decrease sharply with the age of the muscle. 
Consequently the data obtained for the oldest ages are dramatically more weighted in the fitting process.
The shaded regions represent the best fit of the normalized and aggregated data for each muscle at plus and minus one standard deviation. 
We then have $\zeta_{\rm{EDL}} = 0.27 \pm 0.02$ and $\zeta_{\rm{SOL}} = 0.23 \pm 0.02$, which verifies that $r_{\rm{SOL}} < r_{\rm{EDL}}$.
By recognizing that $\frac{\zeta_{\rm{EDL}}}{\zeta_{\rm{SOL}}} = \frac{1 - r_{\rm{SOL}}}{1 - r_{\rm{EDL}}}$, we obtain the condition
$r_{\rm{EDL}} \geq 1 - \frac{\zeta_{\rm{SOL}}}{\zeta_{\rm{EDL}}}$. A strict equality in the latter expression would imply that $r_{\mathrm{SOL}}$ 
is exactly zero, i.e., the soleus muscle is under the Dirichlet boundary condition. Based on the experimental data, we can therefore conclude, 
with $r_{\rm{SOL}} > 0$, that $r_{\rm{EDL}} > 0.14 \pm 0.08$.

%\textbf{Attention, dans le cas ou }$f_{\mathbf{m}}$\textbf{ est vraiment
%petit, alors la caract\'{e}ristique devient intrins\`{e}quement courbe. (cf
%Caplan 1966 figure 8) }
%
%\textbf{(voir aussi isocin\'{e}tique et isoinertiel Tihany 1982 Thorstenson
%1976 pour la relation forme de courbe et pourcentage de fibre, voir aussi
%sargeant 1994 et sargeant dolan 1984)}

\subsubsection{Speed-dependence of the extra heat term}
In his seminal work \cite{Hill1938} Hill identified the three parameters as constants, following the viscoelastic origin of the muscle response proposed
by Gasser and Hill~\cite{Gasser1924}. As indicated in the appendix,
Hill's 1938 model is based on two hypotheses, one of which concerns
the heat released during muscle contraction, which Hill proposed to
formulate as proportional to the rate of contraction, $H=av$, where
$a$ is the so-called \textit{extra heat coefficient} considered to be
constant. Interestingly enough, this dependence, initially ignored by
Hill was considered by Fenn \cite{Fenn1923,Fenn1935,Fenn1924}, and
later reconsidered in experiments by Aubert
\cite{Aubert1956,McMahon1984}, and finally taken up by Hill himself
\cite{Hill1964}. A review of these effects can be found in
\cite{Rall1976,Rall1985}. Indeed, the correction on this term is very
generally small, which very often leads to not detecting it
experimentally. However, if we consider the complete expression
proposed in Eq.~(\ref{a}), we note that the corrective term involves
the resistance to mechanical displacement $R_M$, which corresponds to
the viscosity opposing the displacement of the strands within the
muscle fibers. This results in a corrected expression of type
$H=aI_{M}+R_{M}I_{M}^{2}$. This correction, although small, is in no
way incidental because neglecting it amounts to assimilating, from a
mechanical point of view, the muscle fibre to a generator without
internal resistance, which could therefore lead to an infinite speed
displacement and an equally infinite power production. In practice
this situation is not realized because this resistance is in series
with the feedback resistance $R_{\rm fb}$ which overrides $R_M$ in the
mechanical viscous dissipation, which leads to masking the effect of
$R_M$. From an experimental point of view, by relying on
Eq.~(\ref{a}), it is easy to consider the situations for which this corrective term becomes visible. To do this, it is important that the metabolic intensity, and therefore the rate of contraction, can reach high values. This is only possible in cases where $I_T$ itself is important, i.e. in the presence of a dominant presence of fast fibres. In this case, the hyperbolic form of the response of the force-speed response curve gives way to a much more rectilinear profile, especially at high velocities as previously described. 

The above developments show that it is possible to describe the
process of converting chemical energy into mechanical energy in the
same terms as any thermodynamic machine composed of a working fluid
with a figure of merit, on the one hand, and a device that implements
this working fluid on the other hand. In this case, glucose
degradation remains the central element to which a figure of merit can
be associated. Depending on the coupling conditions to the reservoirs
governed by $R_+$, this degradation can occur quickly or slowly
(glycolytic or oxidative route). More precisely, the kinetic of the
glucose combustion, fast or slow, requires the presence of direct
$(r\approx 0)$, or reduced, $(r\approx 1)$ access to the resource. But
$r$, hence $R_+$, does not account for glucose combustion kinetics
rate itself, 
but the latter is compatible only with certain values of $r$. In other words, there is no use to have an efficient device if it is not correctly fed. It could be argued that in this case the condition of direct access to the resource $(r\approx 0)$ should apply systematically since it would ensure a sufficient energy start for any type of muscle, but this is not the case as for slow fibres it would lead to unnecessarily high basal consumption. Our present model does not claim to explain the detailed metabolic pathways, which have been the subject of much work, but rather to characterize the chemical-mechanical conversion through a systemic approach. If by its fast kinetics, inherited from Dirichlet coupling conditions, the glycolytic pathway allows both force and power production, its high basal consumption is nonetheless incompatible with a prolonged effort. On the other hand, due to its Neumann coupling condition, the oxidative pathway exhibits a strong reduction in basal power, allowing a prolonged muscle activity at low power. Therefore, the glycolytic and oxidative pathways are perfectly adapted to the boundary conditions to which they are subjected.

\begin{figure}[ptbh]
\centering
% Figure-5DANGELO.jpg
\includegraphics[width=0.9\textwidth]{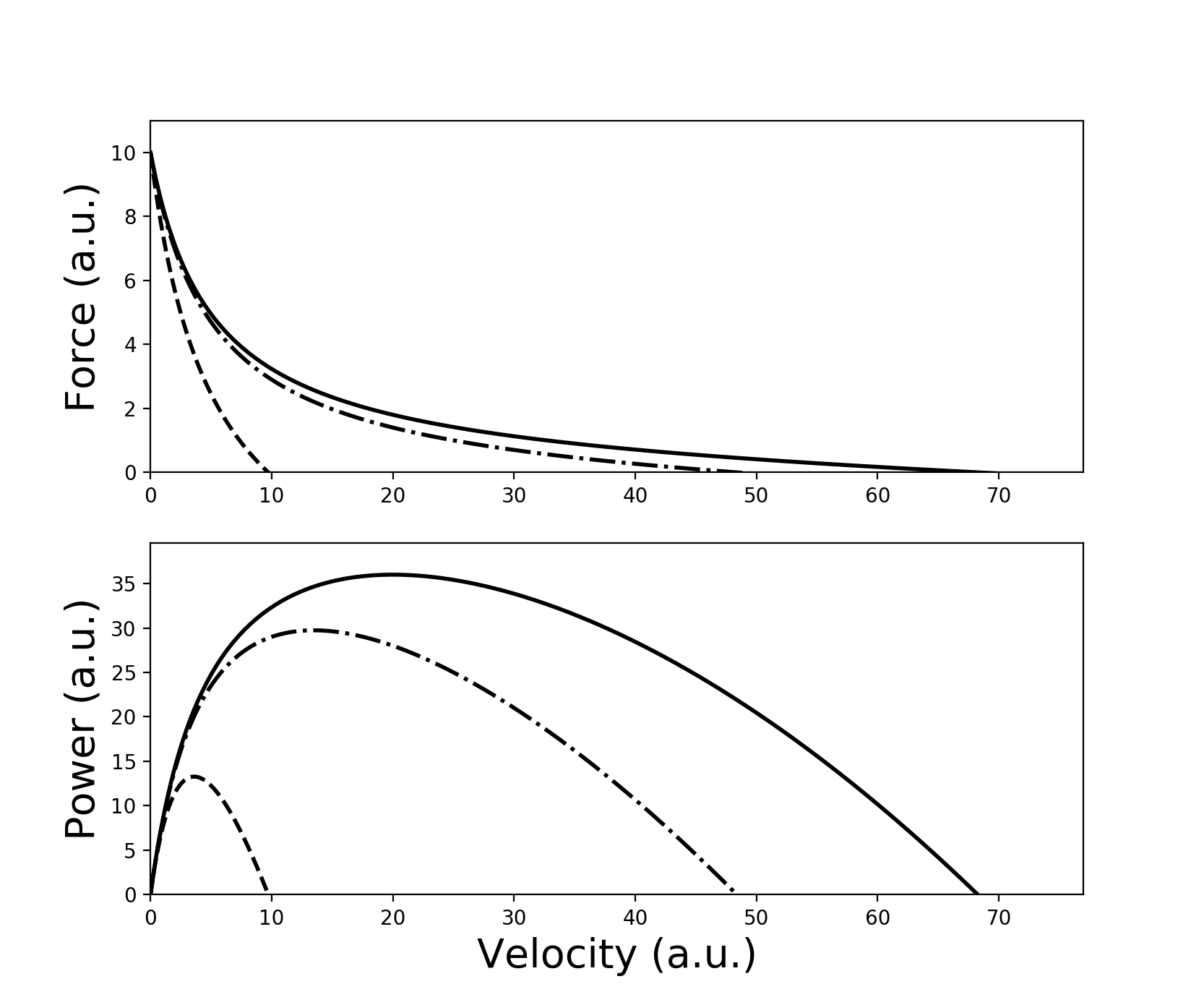}\caption{Force-velocity, and power-velocity, response. The three curves differ only in the boundary conditions given by the values of $r$: $0.05, 0.15, 0.45$ ranging from fast fibres majority to slow fibres majority muscle (see Table \ref{table2} for expressions). The line dot curves give the response vs shortening velocity for constant $a_0$, i.e.,  1938 historical Hill model \cite{Hill1938}. The solid line gives the response for non constant $a$ parameter, as reported by Hill in 1964 \cite{Hill1964,McMahon1984}, and shows reduced performances as predicted with our modelling.}
\label{Hill}%
\end{figure}

As mentioned above, the model only considers short-term efforts and, as such, does not take into account the influence of waste $(R_{-}=0)$. 
This model can be applied to force-velocity response measurements that are mainly performed under these conditions. An extension of the model 
to the case of prolonged efforts is of course possible, but the variation of the parameter $\mu_-$ would have to be accounted for, which, 
for simplicity, has not been the case in this article. In Fig.~\ref{Hill} the force-velocity response is displayed for various ratios of the boundary 
conditions parameter $r$. For a given value of the figure of merit, the variation of $r$ alone is sufficient to produce the slow or 
fast muscle response curves, while the second figure reproduces the shape of the power curves as they were estimated 
from Hill's initial 1938 model and then from his corrected 1963 model, which coincides with our model. 
Similar result can be found in \cite{McMahon1984}. The figure represents the relative influences 
of the composite resistance $R_{H}(I_{M})+R$. In the most classic case $R_{\rm fb}\gg R_M$ and the muscle response follows Hill's initial hyperbolic law. 
Since $R_{M}$ has a very small value, the feedback term $R_{\rm fb}$ drives the overall response and the dissipation appears to be controlled by the boundary 
conditions. Otherwise, if $R_{\rm fb}<R_M$ then the response becomes linear. This case of high mechanical friction $R_M$ would correspond to a 
particularly low figure of merit, which is unlikely in living systems. Although linear, this response cannot be confused with the response of fast muscles, which are 
characterized by high values of $I_T$. In term of Hill parameters, by noting that $F_{\mathrm{iso} }=c/b-a_{0}$ where $a_{0}=R_{\mathrm{fb}}I_{T}$ 
it comes that $R_{H} (I_M)=c/b(b+I_M)$, and hence 

\begin{equation}
P=\left[\frac{c}{I_M+b}-a\right]I_M 
\end{equation}

\noindent so the maximal intensity is given by, $I_{X}=c/a-b$, which, assuming $a\approx a_{0}$, reduces to 

\begin{equation}
I_{X}\approx\frac{F_{\mathrm{iso}}}{R_{\mathrm{fb}}}.
\end{equation} 

\noindent We observe here that, although Hill's hypothesis, $H=av$ is tantamount to totally neglecting mechanical viscosity, 
i.e., $R_M$ in his description of the muscle (which is a daring hypothesis), this has not had any significant 
consequence because the dynamics of the system is limited by $R_{\rm fb}$ which dominates and defines the breaking point $I_X$.

\section{The maximal power principle revisited}
\label{sec5}
In this last section, we show how, within the framework of our
approach, the so-called ``maximum power principle'', frequently invoked in
biology, can be understood as a specific configuration of an
out-of-equilibrium system, and should not be considered as a fundamental
principle. The capacity of out-of-equilibrium systems to absorb energy was discussed by
Lotka \cite{Lotka1921,Lotka1922a,Lotka1922b}, before a proper framework for
nonequilibrium thermodynamics was put on firm grounds. As recently mentioned
by Sciubba \cite{Sciubba2011}, Lotka's assumption referred to the capacity of
a given system to maximize the capture of free energy, which is also the free
energy fraction of the incoming energy. As it is shown below, Lotka's
description can in fact be reconsidered in terms of impedance matching.
However, some misunderstanding of Lotka's analyses gave rise to the so-called
maximal power principle (MPP), which is sometimes considered as the fourth
principle of thermodynamics. But, strictly speaking, up to now, the MPP never
received final rigorous proof
\cite{Beretta2009,Gyamarti1970,Lucia2007,Martyushev2006,Nicolis1977,Zupanovic2010}%
. On the contrary, there is now an increasing consensus to describe
out-of-equilibrium systems not in terms of MPP, but simply using the classical
principles of thermodynamics, now accounting for specific boundary conditions
\cite{Ouerdane2015b}.

We now reconsider this question in the light of the results of our modelling,
using the generic picture of Fig.~(\ref{YCNCA}). Following Lotka, we consider
free energy circulation. A straightforward analysis of the system reveals that the
free energy can be destroyed according to three different processes: i) the
incoming energy quality is degraded by the presence of $R_{+}$; ii) the basal
metabolism is governed by $R_{E}+R_{+}$, which reduces the available free energy
for other activities; iii) the internal dissipation term $R_{M}$ diminishes
the maximal accessible power. To reduce the effects of these dissipation
sources, we may first consider a configuration where $R_{+}=0$, although this
assumption is obviously unrealistic. The equation (\ref{PetFLUX}) then becomes
$P=\alpha I_{M}\left(  \mu_{+}-\mu_{-}\right)  -R_{M}I_{M}^{2}$ and this means
the animal activity would be only limited by its internal dissipation
$R_{M}I_{M}^{2}$. In addition, the basal metabolism term $B=\frac{\mu_{+}%
-\mu_{-}}{R_{E}+R_{+}}$ increases to its maximal value. In other words the gain obtained by easier
access to the resources, stemming from the assumption $R_{+}=0$, is
counterbalanced by an increase of the basal metabolism. Such a simple extremal
free energy analysis is far from being sufficient to explain the optimal
physiological point, if any.

Since the condition $R_{+}=0$ is clearly unrealistic, we now turn to the two
other cases with $R_{+}\neq0$. According to the process ii) above, a reduction
in the basal metabolism would improve the free energy proportion. Then, in the
configuration where $R_{+}$ is small and $R_{E}/R_{+}\ll1$, Eq.~(\ref{PetFLUX}%
) now gives $P \approx\alpha I_{M}\frac{R_{E}}{R_{+}}\left(  \mu_{+}-\mu
_{-}\right)  -R_{M}\left(  \frac{\alpha^{2}R_{E}}{R_{M}}\mu_{-}+1\right)
I_{M}^{2}$ which means that the available power is vanishingly small, a
totally useless configuration. It then becomes obvious that the reduction of
the basal metabolism under a certain threshold is not desirable, unless it is
imposed by external constraints, such as (and often mainly) food resources. 
In this respect, as previously described, 
slow and fast fibers strategies represent two opposite configurations. 
In short, in order to reach an optimum, a trade-off between the relative values
of $R_{E}$ and $R_{+}$ is unavoidable. Clearly, the implicit variational
principle for an free energy maximum is necessary but not sufficient, and the
feedback induced by the boundary conditions leads to a distribution of the
free energy degradation locations \cite{Goupil2016}. The final distribution depends 
on the degrees of freedom of each system component, and sub-optimal configurations
may exist due to the absence of any degree of freedom on the local level. From
a statistical point of view, it is clear that a complex system, composed of a
network of diverse engines, may present different (sub)optimal configurations.
There exists a constrained matching of the global impedance of the
system, allowing an ``optimal'', but not necessary maximal, free energy flow.

Let us finally discuss the question of the extremal principle, by deriving an
impedance expression. The energy flux is approximately $\Phi\approx{\frac{\mu_{+}-\mu_{-}}{R_{E}+R_{\Sigma}}}$,
where $R_{\Sigma}=R_{+}+R_{-}$, and the chemical potential difference
$\Delta\mu=\mu_{+}-\mu_{-}$ directly governs the output power. Note that
strictly speaking $\Delta\mu=0$ correspond to the \textit{dead body
condition}. Using the argument of the minimization of the degradation of the
free energy, we expect a minimization of $R_{\Sigma}$. In addition, if the ratio
$R_{E}/R_{\Sigma}\rightarrow0$, we see that $\Delta\mu_{M}\rightarrow0$ and
both the output power and the efficiency vanish. Now, if we consider the
situation where $R_{E}/R_{\Sigma}\rightarrow\infty$, the efficiency becomes
maximal, but the power again vanishes. In both cases, we experiment a
\textit{minimal power principle, with maximal or minimal entropy production}.
If we now consider $R_{\Sigma}=0$, then the value of $R_{E}$ does not matter
any more, and the output power will be maximal, possibly diverging if $R$
decreases. This configuration sounds very similar to the so-called MPP
configuration, and actually is, but, as said above, the boundary condition
$R_{\Sigma}=0$ is \textit{not} realistic. If we finally consider that
$R_{\Sigma}$ is small but non zero, then the non vanishing $\Delta\mu$
condition imposes a non-zero value for $R_{E}$. Using elementary algebra, the
optimal configuration is found to be $R_{E}=R_{\Sigma}$, where the output
power is now maximal. If $R_{E}>R_{\Sigma}$ the process favors the energy
conversion efficiency at the expense of power; conversely, if $R_{E}%
<R_{\Sigma}$ the process favors the output power at the expense of efficiency.
Therefore, there is no extremal principle for such out-of-equilibrium systems,
but there exists a perfectly definite configuration, specified by the actual
conditions imposed at the system boundaries.
%}

\section{Concluding remarks}

\label{sec6} Thermodynamics of biological systems entails a rather wide range
of problems and models, and remains a very active and timely field of research
\cite{Roach2018}. In this article, our focus was on the development of a
theoretical framework based on a generalized meta-formulation of linear
non-equilibrium thermodynamics to characterize the production of metabolic
energy and its use under muscular effort. We show that the feedback effects
in biological systems permit an analysis based on a local linearized model
\cite{Goupil2016}, and starting from the first and second principles of
thermodynamics, considered in the context of a locally linear
out-of-equilibrium response, we proposed a model of the thermodynamic response
of an organism under effort, that lies on a reduced (effective) set of
physical quantities: the feedback resistance $R_{\mathrm{fb}}$, the basal
power $B$, the isometric force $F_{\mathrm{iso}}$ and the metabolic figure of
merit $Z\mu$, which we derived from the generalized thermoelastic and
transport coefficients. To actually test the
validity of this approach, we used Hill's model as a touchstone in the sense
that through the analysis of Hill's phenomenological equation we could first establish a 
direct correspondence with the thermodynamic model and also specify the thermodynamical meaning 
of Hill's parameters $a$, $b$, and $c$, including the non-constant behaviour of $a$.

Hill's phenomenological model derives from calorimetry measurement data
\cite{Hill1938} and, as such, the force-velocity response formula,
Eq.~(\ref{H}), is essentially empirical. In the present work, we made it
a principles-based model and we reconsidered many of the questions and debates 
that have marked its history, some of which finding here a common framework.
The main finding of our approach is the active
output impedance, that explains the velocity dependence of Hill's coefficient
$a$ (the so-called \textquotedblleft extra heat\textquotedblright). This
dependence stems from feedback only. There is no fundamental reason why the
active impedance would be simply resistive (i.e. a real number); as a matter
of fact, it is a complex quantity with a non-zero imaginary part. The standard
models proposed since Hill's seminal work contain a complex impedance, but it
is a passive one. Although experimental measurements of muscle response
exhibit an overshoot in the frequency space, this overshoot cannot be
explained assuming a passive impedance model. Note that it has been
established \cite{Goupil2016} that this type of response may exist if the
figure of merit is large, and at the condition that the boundary conditions
are of the mixed type, which implies feedback effects and hence an increase of
the order of the transfer function of the looped response with respect to the
open loop. 

\section*{Acknowledgments}
The authors wish to thank Henri Benisty for interesting discussions and precious comments on draft
versions of the paper.

\appendix
\section{Hill's calorimetry developpement}
\label{annexe2}

In his seminal article Hill presented an extensive experimentation of the
calorimetric response of muscle under contraction. The total energy rate
budget was summarized into four terms as,%
\[
E=A+H+W
\]

\noindent where $E$ is the total energy rate released, $A$ is the maintenance energy
rate, $W=Fv$ is the rate of work, and $H$ the shortening heat rate. Hill
identified the term $H+W$ as the rate of ``extra'' energy induced by the
contribution of additional chemical reactions during contraction. The term $H$
was clearly identified as a chemical parameter, which Hill assumed to be related to the 
difference between the isometric and actual force: 

\begin{equation}
\label{HW}H+W=b(F_{0}-F)
\end{equation}

\noindent In addition, Hill bridged calorimetry and mechanics, considering that 

\begin{equation}
\label{H}H=av
\end{equation}

\noindent And combining Eq.~(\ref{HW}) and Eq.~(\ref{H}) he derived the following 
muscle force-velocity relationship:  

\[
(F+a)\left(v+b\right) = b(F_{0}+a)
\]

\noindent which is the classical Hill's equation. It is important to note that by
writing $H=av$, Hill assumed that the presence of chemical reactions results
in a dissipation term that varies linearly with velocity, which is a strong approximation
but sufficient to explain most, but not all, of the experimental results. 

\section*{References}

\bibliography{sample}

\newpage

%\begin{figure}[ptbh]
%\centering
%\includegraphics[width=0.3\textwidth]{Figure-7DANGELO.jpg}\caption{Schematic
%view of thermodynamic engine under mixed boundary conditions.}%
%\label{MPP}%
%\end{figure}

\end{document}